\documentclass[12pt]{article}
\usepackage{fullpage,citesort,epsfig,psfrag,graphics,amsbsy,amssymb}
\usepackage{caption}
\newcommand{\phys}{{\rm phys}}
\newcommand{\phyk}{\ket{\rm phys}}

\newcommand{\beq}{\begin{equation}}
\newcommand{\eeq}{\end{equation}}
\newcommand{\bea}{\begin{eqnarray}}
\newcommand{\eea}{\end{eqnarray}}

\newcommand{\be}{\begin{equation}}
\newcommand{\ee}{\end{equation}}
\newcommand{\bq}{\begin{eqnarray}}
\newcommand{\eq}{\end{eqnarray}}
\newcommand{\ket}[1]{|#1\rangle}
\newcommand{\bra}[1]{\langle#1|}

\def\math{\mathsurround=0pt }
\def\leftrightarrowfill{$\math \mathord\leftarrow \mkern-6mu 
 \cleaders\hbox{$\mkern-2mu \mathord- \mkern-2mu$}\hfill
 \mkern-6mu \mathord\rightarrow$}
\def\overleftrightarrow#1{\vbox{\ialign{##\crcr
     \leftrightarrowfill\crcr\noalign{\kern-1pt\nointerlineskip}
     $\hfil\displaystyle{#1}\hfil$\crcr}}}

\newcommand{\VEV}[1]{\langle#1\rangle}

\let\l=\lambda

 \def\bd{\begin{document}} \def\ed{\end{document}}
\def\ds{\documentstyle} \let\fr=\frac \let\bl=\bigl \let\br=\bigr
\let\Br=\Bigr \let\Bl=\Bigl
\let\bm=\bibitem
\let\na=\nabla
\let\pa=\partial \let\ov=\overline
\def\ft#1#2{{\textstyle{{\scriptstyle #1}\over {\scriptstyle #2}}}}
\def\fft#1#2{{#1 \over #2}}
\def\vp{\varphi}
\def\sst#1{{\scriptscriptstyle #1}}
\def\oneone{\rlap 1\mkern4mu{\rm l}}
\def\td{\tilde}
\def\wtd{\widetilde}
\def\dalemb#1#2{{\vbox{\hrule height .#2pt
        \hbox{\vrule width.#2pt height#1pt \kern#1pt
                \vrule width.#2pt}
        \hrule height.#2pt}}}
\def\square{\mathord{\dalemb{6.8}{7}\hbox{\hskip1pt}}}
\def\wtd{\widetilde}
\def\R{\rlap{\rm I}\mkern3mu{\rm R}}
\def\im{{\rm i}}
\def\tilg{\tilde{g}}
\def\tilF{\tilde{F}}
\def\tilA{\tilde{A}}
\def\varf{\varphi}
\def\tilf{\tilde{\phi}}
\def\tilh{\tilde{h}}
\def\rme{{\rm e}}
\def\ep{\epsilon}
\def\0{{(0)}}
\def\9{{(9)}}
\def\8{{(8)}}
\def\7{{(7)}}
\def\6{{(6)}}
\def\5{{(5)}}
\def\4{{(4)}}
\def\3{{(3)}}
\def\2{{(2)}}
\def\1{{(1)}}
\newcommand{\trace}{{\rm Tr}}
\newcommand{\ub}{\overline{U}}
\newcommand{\vb}{\overline{V}}
\newcommand{\uh}{\widehat{U}}
\newcommand{\vh}{\widehat{V}}
\newcommand{\ubh}{\overline{\widehat{U}}}
\newcommand{\vbh}{\overline{\widehat{V}}}
\newcommand{\lb}{\bar{\l}}
\newcommand{\Fb}{\overline{F}}
\newcommand{\Fh}{\widehat{F}}
\newcommand{\Fbh}{\overline{\widehat{F}}}
\newcommand{\Ab}{\overline{A}}
\newcommand{\Ah}{\widehat{A}}
\newcommand{\Abh}{\overline{\widehat{A}}}
\newcommand{\Gb}{\overline{G}}
\newcommand{\Gh}{\widehat{G}}
\newcommand{\Gbh}{\overline{\widehat{G}}}
\newcommand{\Pb}{\overline{P}}
\newcommand{\Ph}{\widehat{P}}
\newcommand{\Pbh}{\overline{\widehat{P}}}
\newcommand{\Qb}{\overline{Q}}
\newcommand{\Qh}{\widehat{Q}}
\newcommand{\Qbh}{\overline{\widehat{Q}}}
\newcommand{\Bb}{\overline{B}}
\newcommand{\Bh}{\widehat{B}}
\newcommand{\Bbh}{\overline{\widehat{B}}}
\newcommand{\fhns}{\hat{F}^{\rm (NS)}}
\newcommand{\fhrr}{\hat{F}^{\rm (RR)}}
\newcommand{\ahns}{\hat{A}^{\rm (NS)}}
\newcommand{\ahrr}{\hat{A}^{\rm (RR)}}
\newcommand{\hhrr}{\hat{H}^{\rm (RR)}}
\newcommand{\hchi}{\hat{\chi}}
\newcommand{\hphi}{\hat{\phi}}
\newcommand{\htau}{\hat{\tau}}
\newcommand{\cG}{{\cal G}}
\newcommand{\cGb}{\overline{{\cal G}}}
\newcommand{\cH}{{\cal H}}
\newcommand{\cP}{{\cal P}}
\newcommand{\cPb}{\overline{{\cal P}}}
\newcommand{\cQ}{{\cal Q}}
\newcommand{\cQb}{\overline{{\cal Q}}}
\newcommand{\cM}{{\cal M}}
\newcommand{\cN}{{\cal N}}
\newcommand{\cO}{{\cal O}}
\newcommand{\cD}{{\cal D}}
\newcommand{\cL}{{\cal L}}

\newcommand{\vpp}{\mbox{$\langle{\scriptstyle++}\rangle$}}
\newcommand{\vmp}{\mbox{$\langle{\scriptstyle-+}\rangle$}}
\newcommand{\vppp}{\mbox{$\langle{\scriptstyle+++}\rangle$}}
\newcommand{\vmpp}{\mbox{$\langle{\scriptstyle-++}\rangle$}}
\newcommand{\vpmp}{\mbox{$\langle{\scriptstyle+-+}\rangle$}}

\begin{document}
\setlength{\captionmargin}{36pt}
\begin{titlepage}
\phantom{.}

\vskip 3cm
\begin{center}
\begin{large}
{\bf Null Physical States in String Models}
\end{large}

\vskip 2cm
{\large 
Charles B. Thorn\footnote{E-mail  address: {\tt thorn@phys.ufl.edu}}
}
\vskip0.8cm
{\it Institute for Fundamental Theory,\\
Department of Physics, University of Florida,
Gainesville FL 32611}


\vskip 1.0cm
\end{center}

\begin{abstract}
\noindent This note is a brief addendum to my
article Nucl.\ Phys.\ B {\bf 864} (2012) 285,
[arXiv: 1110.5510], which discusses the
noghost theorem in Ramond sectors of string models. 
In this addendum we derive additional information about the structure of 
null physical states in the Ramond-Neveu-Schwarz model.
\end{abstract}
\vfill
\end{titlepage}
\section{Introduction}
The structure of null states lies at the heart of the proofs of the 
noghost theorems for string models. This is true in the 
original versions 
\cite{goddardt,brower,corrigang}, their improved
versions \cite{thorndetailed,thornimpng}, and also in the modern BRST-based
versions \cite{katoogawa,thorndetailed,thornimpng}.
In all cases,  one shows that in the critical dimension
($D=26$ for the bosonic models and $D=10$ for strings
based on Ramond-Neveu-Schwarz models \cite{rns})
all physical states, i.e. all states that couple to physical
on-shell processes, can be expressed as
\bea
\ket{\rm phys}&=&\ket{\rm null}+\ket{T} 
\eea
where the transverse states $\ket{T}$ span a positive definite
subspace of the physical states, and 
the null states have zero overlap with themselves
and with all physical states. The state spaces of the
superstring and other derivative
string models lie within the state spaces of these parent 
critical string models, and so are also covered by these
theorems.

While some properties of the null states are derived in the
course of proving the noghost theorem, there are more
detailed facts about them that require further
argumentation to establish. For example, in the appendix of
my paper \cite{thorndetailed}, which streamlined the original
Goddard-Thorn proof \cite{goddardt}, I proved that all
on-shell ($[L_0-1]\ket{\rm null}=0$) null states in the bosonic model 
can be expressed in the form
\bea
\ket{\rm null}&=&L_{-1}\ket{\rm phys}_1+\left(L_{-2}+{3\over2}L^2_{-1}\right)
\ket{\rm phys}_2.
\label{nullbosonic}
\eea
where $L_n$ are the generators of the Virasoro algebra. The
states $\ket{\rm phys}_{1,2}$ are annihilated by all $L_n$ with
$n>0$. In the language of conformal field theory this means they
are primary states, and the above equation states that all null 
states are either of two particular descendants of primary
states. Such a classification of null states has proved
useful in some investigations, for example Witten's
recent treatise on superstring perturbation theory \cite{wittenrevisited}.

Analogous facts are true of the non-bosonic string models.
We will use the notation of the original papers: the super-Virasoro
generators will be denoted $F_n,L_n$ in Ramond sectors and
$G_r,L_n$ in Neveu-Schwarz sectors. Indices $m,n$ will run over all
integers, and indices $r,s$ will run over all half-odd integers.
In Neveu-Schwarz sectors of such models the analog of (\ref{nullbosonic}) 
reads
\bea
\ket{\rm null}&=&G_{-1/2}\ket{\rm phys}_1+\left(G_{-3/2}+{1\over2}
G_{-1/2}L_{-1}\right)\ket{\rm phys}_2.
\label{nullns}
\eea
the proof of which is a straightforward generalization of the one
in the appendix of \cite{thorndetailed}. However
the corresponding generalization to Ramond sectors 
\bea
\ket{\rm null}&=&F_0F_{-1}\phyk_1+ F_0L_{-1}\phyk_2
\label{nullramond}
\eea
is less straightforward because of zero mode
complications, so we devote the remainder of this short note to
explaining it. The corresponding argument for Neveu-Schwarz
sectors is completely parallel but with the absence of zero-mode
complications.   
In the following section 2 I recall some results from \cite{thornimpng}
that are necessary to complete the proof of the validity of
(\ref{nullramond}), after which I complete the proof of (\ref{nullramond}) 
in section 3.
\section{An ordered basis}
Here we gather results from \cite{thornimpng} that we will need 
later.
The super-Virasoro algebra in $D$ spacetime dimensions reads:
\bea
 {}[L_n, L_m]&=&(n-m)L_{n+m}+{D\over8}n^3\delta_{n,-m}\\
 {}[L_n, F_m]&=&\left({n\over2}-m\right)F_{n+m}\\
 {}\{F_n, F_m\}&=&2L_{n+m}+{D\over2}n^2\delta_{n,-m}\;.
\eea
In the following we will always assume the critical dimension
$D=10$. A lightlike vector $k^\mu$ is chosen to define 
$D_n=k\cdot d_n$ and $K_n=k\cdot a_n$, where $d^\mu_n$ and $a^\mu_n$ are
the fermionic and bosonic modes respectively of the Ramond sector.
We will always work in the eigenspace of energy-momentum with
value $p^\mu$, and we normalize $k$ so that $K_0=1$.
Then
\bea
 {}[L_n,K_m]&=&-mK_{m+n},\qquad [L_n,D_m]=-\left(m+{n\over2}\right)D_{m+n}\\
 {}[F_n,K_m]&=&-mD_{m+n},\qquad \{F_n,D_m\}=K_{m+n}\\
 {}[K_n,K_m]&=&0,\qquad\{D_n,D_m\}=0,\qquad [K_n,D_m]=0\;.
\eea
The physical states are annihilated by all $L_n$, $F_n$ with $n>0$. 
The transverse states $\ket{T}$ are physical states that in addition
are annihilated by $K_n$
for $n>0$ and by $D_n$ for $n\geq0$. Any two transverse states have vanishing
inner product $\VEV{T|T^\prime}=0$, and nonzero inner products require the insertion of
an $F_0$ factor:
\bea
\bra{T}F_0\ket{T^\prime}&\neq&0\; .
\eea
Defining the norm with this inner product, the transverse states
have nonnegative norm, relative to an overall constant factor. 

In \cite{thornimpng} we established that the basis set of the whole 
Ramond sector state space,
\bea
\ket{\{f\}\{\lambda\},\{d\}\{\kappa\}}
=F_0^{f_0}F_{-1}^{f_1}L_{-1}^{\lambda_1}\cdots 
F_{-l}^{f_l}L_{-l}^{\lambda_l}
D_{-1}^{d_1}\cdots D_{-k}^{d_k}
K_{-1}^{\kappa_1}\cdots K_{-k}^{\kappa_k}\ket{T}
\label{lkbasis}
\eea
where $\ket{T}$ are arbitrary transverse states, is linearly
independent. The labels $\{\lambda\}$ and $\{\kappa\}$ are 
bosonic partitions of 
two nonnegative integers. Similarly $\{f\}$ and $\{d\}$
are fermionic partitions of two nonnegative integers. 
Fermionic simply means that each $f_i$ and $d_i$
assumes only the values 0 or 1. 
A conjugate (or ``dual'') to each
element (\ref{lkbasis}) is defined by
\bea
\ket{\{f\}\{\lambda\},\{d\}\{\kappa\},C}
=F_0^{1-f_0}F_{-1}^{d_1}L_{-1}^{\kappa_1}\cdots 
F_{-k}^{d_k}L_{-k}^{\kappa_k}
D_{-1}^{f_1}\cdots D_{-l}^{f_l}
K_{-1}^{\lambda_1}\cdots K_{-l}^{\lambda_l}\ket{T}\; .
\label{conjbasis}
\eea
The inner product of each basis element
with its conjugate is not zero.

We also defined an ordering of this basis by ordering the
partitions $\{f\},\{\lambda\}$, according to $(\{f,\lambda\})<(\{f^\prime,
\lambda^\prime\})$ if the first nonzero entry of the sequence
\bea
\sum_{i}i(f_i-f_i^\prime+\lambda_i-\lambda_i^\prime),\quad f_0-f_0^\prime, 
\quad f_1-f_1^\prime,\quad \lambda_1-\lambda_1^\prime,\quad f_2-f_2^\prime,
\quad
\cdots
\label{ordering}
\eea
is positive. A similar ordering is defined for the other pair of 
partition labels $\{d\kappa\}$.
Then we order the entire basis according to $(\{f\lambda\},\{d\kappa\})
<(\{f^\prime\lambda^\prime\},\{d^\prime\kappa^\prime\})$
if $\{f\lambda\}<\{f^\prime\lambda^\prime\}$ or if 
$\{f\lambda\}=\{f^\prime\lambda^\prime\}$ and
$\{d\kappa\}>\{d^\prime\kappa^\prime\}$.
With this ordering we then quote a crucial result of \cite{thornimpng},
which will also be needed in the following section: 
\bea
\VEV{\{f\}\{\lambda\},\{d\}\{\kappa\},C|\{f^\prime\}\{\lambda^\prime\},
\{d^\prime\}\{\kappa^\prime\}}=0,\qquad {\rm if}\quad(\{f\lambda\},\{d\kappa\})<
(\{f^\prime\lambda^\prime\},\{d^\prime\kappa^\prime\})\; ,
\label{triangularity}
\eea
which is to say that the corresponding matrix of inner products
is lower triangular. 
\section{Null States}
We first enumerate all physical states, those annihilated by 
$L_n$ and $F_n$ for all $n>0$, on and off shell. They are spanned by the basis
\bea
\ket{\{f\}\{\lambda\},\phys}
&=&F_0^{f_0}F_{-1}^{f_1}L_{-1}^{\lambda_1}\cdots 
F_{-l}^{f_l}L_{-l}^{\lambda_l}\ket{T}+{\rm Terms~with}~\{d,\kappa\}\neq0.
\label{physbasis}
\eea
For each fixed $\{f,\lambda\}$ the unlisted terms are uniquely
determined.\footnote{To see this one applies in turn, in
the order highest to lowest according to (\ref{ordering}), the monomials
$L_l^{\lambda_l}F_l^{f_l}\cdots L_1^{\lambda_1}F_1^{f_1}$ to a general linear 
combination of the basis states (\ref{lkbasis}). Then because of
the triangularity (\ref{triangularity}) the action of the monomial picks out one by one 
the terms with $\{d,\kappa\}\neq0$ which produce a term with
$\{d,\kappa\}=0$. This unique term can only be cancelled by
states produced by the action of the monomial on terms
with $\{d,\kappa\}=0$. Thus all physical states must
have at least one term with $\{d,\kappa\}=0$, and further the
structure of the states (\ref{physbasis}) is uniquely determined.}  
The first term, which completely determines each
such physical state will be called the leading term. In the following we will 
frequently be working with that term alone with all the others implied.

The following are on-shell null states ($L_0=0$):
\bea
F_0F_{-1}\phyk_1,\qquad F_0L_{-1}\phyk_2
\label{nullstates}
\eea
as can be seen by a short direct calculation. The on-shell condition
means that the $L_0$
eigenvalues of $\phyk_{1,2}$ are always -1.
In the following we show that these are all of the on-shell null
states. We can enumerate the states $\phyk_1,\phyk_2$ via the basis
(\ref{physbasis}), but there are linear dependences among
the states (\ref{nullstates}) in that labeling.
First of all, from $F_0^2=L_0=0$ on-shell and the superconformal algebra,
we have the proportionalities
\bea
F_0F_{-1}F_0F_{-1}&\propto& F_0L_{-1}F_{-1},\quad F_0F_{-1}F_0
\propto F_0L_{-1}\\
 F_0L_{-1}F_0F_{-1}&\propto& F_0F_{-1}F_{-1},\quad F_0L_{-1}F_0
\propto F_0F_{-1}
\eea
Thus those basis states contributing to $\phyk_1,\phyk_2$ with
leading terms with $f_0=1$ give the same contribution 
to the null state as those with $f_0=0$. (Recall that the Null states
are physical and that the contribution of each basis element
is uniquely fixed by the leading term).

Each on-shell basis element has $f_0=1$.
Substituting in turn each of the basis elements with $f_0=f_1=0$ for $\phyk_1$,
we see that we will generate all on-shell physical basis elements
with $f_0=f_1=1$. To obtain the states with $f_1=0$, we examine
the second class of states. Substituting in turn each of the 
basis elements with 
$f_0=f_1=0$ for $\phyk_2$,
we produce all on-shell physical basis elements
with $f_0=1,f_1=0,\lambda_1\geq1$. 

We proceed step by step. Next substitute each of the basis elements with 
$f_0=0,f_1=1,\lambda_1=0$ for $\phyk_2$, we find a leading term that starts 
with $F_0L_{-1}F_{-1}\cdots$, which is out of canonical order.
We then use the algebra to rearrange
\bea
F_0L_{-1}F_{-1}&=&F_0F_{-1}L_{-1}+{1\over2}F_0F_{-2}
\eea
which puts the factors in canonical order. We can subtract a null
state of the first type to cancel away the physical state associated
with the the leading term from the first term on the right from the
null state we just formed, so that what remains is all of the
physical states associated with a leading term with $f_0=1,f_1=\lambda_1=0$
and $f_2=1$. To find the states with  $f_0=1,f_1=\lambda_1=f_2=0$,
we substitute basis elements with $f_1=1,\lambda_1=f_2=0$ for
$\phyk_1$. Then, because $F_{-1}^2=L_{-2}$,
the leading term of the resulting null state has $f_0=1,f_1=\lambda_1=f_2=0$,
and $\lambda_2\geq1$. 

We next look for states with $f_0=1,f_1=\lambda_1=f_2=\lambda_2=0$.
First substitute the physical states with 
leading term for which $f_0=0,f_1=1,\lambda_1=1,f_2=\lambda_2=0$
for $\phyk_2$. Then rearrange
\bea
L_{-1}F_{-1}L_{-1}&=&{1\over2}F_{-2}L_{-1}+F_{-1}L_{-1}^2
=-{1\over2}\cdot{3\over2}F_{-3}+{1\over2}L_{-1}F_{-2}+F_{-1}L_{-1}^2
\eea
which puts all the operators in canonical order. The contributions
from the second two terms will produce the leading terms of
null states previously accounted for, so they can be cancelled away
leaving all the null physical states with leading terms with
$f_0=1,f_1=\lambda_1=f_2=\lambda_2=0$ and $f_3=1$, i.e. of the form
\bea
F_0F_{-3}L_{-3}^{\lambda_3}\cdots\ket{T}.
\eea
To get states with $f_3=0$, we substitute basis elements 
with $f_1=1,\lambda_1=1,f_2=\lambda_2=0$ for
$\phyk_1$. Then, using $F_{-1}^2=L_{-2}$, we rearrange 
\bea
L_{-2}L_{-1}=-L_{-3}+L_{-1}L_{-2}
\eea
The second term produces null stated already accounted for.
Subtracting them leaves all null physical states with
$f_0=1, f_1=\lambda_1=f_2=\lambda_{2}=f_3=0$ and $\lambda_3\geq1$. 

From here on we just continue this process recursively. At the 
$n$th step we first substitute the physical states with 
leading term for which $f_0=0,f_1=1,\lambda_1=n,f_2=\lambda_2
=f_3=\lambda_3=\cdots=f_{n+1}=\lambda_{n+1}=0$ for $\phyk_2$.
Then rearrange
\bea
L_{-1}F_{-1}L_{-1}^n&=&F_{-1}L_{-1}^{n+1}+{1\over2}F_{-2}L_{-1}^n\nonumber\\
&=&F_{-1}L_{-1}^{n+1}+{1\over2}L_{-1}\sum_{k=0}^{n-1}
k!{-3/2\choose k}{n\choose k}L_{-1}^{n-k-1}F_{-(k+2)}\nonumber\\
&&\qquad+(-)^n{1\over2}\cdot{3\over2}\cdots{2n+1\over2}F_{-2-n}
\eea
all terms but the last term on the right produce the leading terms
of null states previously accounted for, so they can be cancelled
away leaving the null states with $f_0=1,f_1=\lambda_1=f_2=\cdots
f_{n+1}=\lambda_{n+1}=0$ and $f_{n+2}=1$.

The second part of the $n$th step is to substitute physical states
with leading terms for which $f_0=0,f_1=1,\lambda_1=n,f_2=\lambda_2
=f_3=\lambda_3=\cdots=f_{n+1}=\lambda_{n+1}=f_{n+2}=0$ for $\phyk_1$.
Then, using $F_{-1}^2=L_{-2}$, we rearrange
\bea
L_{-2}L_{-1}^n&=&\sum_{k=0}^{n-1}(-)^kk!{n\choose k}L_{-1}^{n-k}L_{-2-k}
+(-)^nn!L_{-n-2}\eea
so all terms are in canonical order. All terms but the last term 
on the right produce the leading terms
of null states previously accounted for, so they can be cancelled
away leaving the null states with $f_0=1,f_1=\lambda_1=f_2=\cdots
f_{n+1}=\lambda_{n+1}=f_{n+2}=0$ and $\lambda_{n+2}\geq1$.

Induction on $n$ then shows that every element in the basis of
on-shell ($L_0=0$) physical states (\ref{physbasis}) 
with $f_0=1$ are contained in the list of
null states (\ref{nullstates}), which is thus complete.
\vskip8pt
\noindent\underline{Acknowledgments}: 
This research was supported in part by the Department
of Energy under Grant No. DE-FG02-97ER-41029.

\end{document}